\documentclass[12pt]{iopart}
\usepackage{graphicx}
\usepackage{amsfonts}

\begin{document}

\title[]{Modulation of surface plasmon coupling-in by one-dimensional surface corrugation}

\author{F L\'opez-Tejeira$^1$, Sergio G Rodrigo$^1$, L
Mart\'{\i}n-Moreno$^1$, F J Garc\'{\i}a-Vidal$^2$, E Devaux$^3$, J
Dintinger$^3$\footnote{Present address: Nanophotonics and
Metrology Laboratory, Swiss Federal Institute of Technology
Lausanne (EPFL), CH-1015 Lausanne, Switzerland}, T W Ebbesen$^3$,
J R Krenn$^4$, I P Radko$^5$, S I Bozhevolnyi$^5$, M U
Gonz\'alez$^6$\footnote{Present address: ICFO-Institut de
Ci\`{e}ncies Fot\`{o}niques, E-08860 Castelldefels, Spain}, J C
Weeber$^6$ and A Dereux$^6$}

\address{$^1$ Departamento de F\'{\i}sica de la Materia Condensada-ICMA,
Universidad de Zaragoza, E-50009 Zaragoza, Spain}
\address{$^2$ Departamento de F\'{\i}sica Te\'orica de la Materia
Condensada, Universidad Aut\'onoma de Madrid, E-28049 Madrid,
Spain}
\address{$^3$ Laboratoire de Nanostructures, ISIS, Universit\'e
Louis Pasteur, F-67000 Strasbourg, France}
\address{$^4$ Institute of
Physics, Karl Franzens University, A-8010 Graz, Austria}
\address{$^5$ Department of Physics and
Nanotechnology, Aalborg University, DK-9220 Aalborg, Denmark}
\address{$^6$ Laboratoire de Physique de l'Universit\'e de
Bourgogne, UMR CNRS 5027, F-21078 Dijon, France}
\ead{lmm@unizar.es}

\begin{abstract}
Surface plasmon-polaritons have recently attracted renewed
interest in the scientific community for their potential in
sub-wavelength optics, light generation and non-destructive
sensing. Given that they cannot be directly excited by freely
propagating light due to their intrinsic binding to the metal
surface, the light-plasmon coupling efficiency becomes of crucial
importance for the success of any plasmonic device. Here we
present a comprehensive study on the modulation (enhancement or
suppression) of such coupling efficiency by means of
one-dimensional surface corrugation. Our approach is based on
simple wave interference and enables us to make quantitative
predictions which have been experimentally confirmed at both the
near infra-red and telecom ranges.
\end{abstract}
\pacs{73.20.Mf,78.67.-n,41.20.Jb} \submitto{\NJP} \maketitle

\section{Introduction}
\label{intro}

Surface plasmon-polaritons (SPPs) are electromagnetic modes
originated from the interaction between light and mobile surface
charges, typically the conduction electrons in metals
\cite{Raetherbook}. Because of the so-called ``excess of
momentum'' with respect to light of the same frequency, SPPs
cannot propagate away from a planar surface and are thus bound to
and guided by it. As a consequence of such binding, SPP modes can
be laterally confined below the diffraction limit, which has
raised the prospect of SPP-based photonic circuits
\cite{Barnes2003,Maier05,Ozbay2006}. To build up this kind of
circuits one would require a variety of components in which
incident light would be first converted in SPPs, propagating and
interacting with different devices before being recovered as
freely propagating light. Hence, a great deal of attention has
been recently devoted to the creation of optical elements for SPPs
\cite{Weeber01,Krenn03,Weeber04,JGRapl06,Bozhecpp06,Gonzalez2006},
as well as to the efficient coupling of freely-propagating light
into and out of them. This latter issue constitutes the
fundamental bottleneck that must be overcome in order to fully
exploit the potential of SPPs, given that established techniques
for SPP generation (which make use of prism
\cite{Otto1968,Lamprecht01}, grating \cite{Ritchie1968} or
nanodefect \cite{Ditlbacher02a} coupling) require that the
system's size be well out of the sub-wavelength scale in order to
obtain a neat SPP signal. On the other hand, $p$-polarized
back-side illumination of sub-wavelength apertures in optically
thick metal films
\cite{Sonnichsen00,Devaux2003,Yin2004,Popov2005,Agrawal2005,Chang2005,Lalanne2005,Lalanne07prl}
prevents both damping and signal blinding but it does not ensure a
unique propagation direction for the generated SPPs.

In a previous work \cite{flt07np}, we proposed a novel back-side
slit-illumination method based on drilling a periodic array of
indentations at one side of the slit. It was demonstrated that the
SPP beam emerging from the slit to its corrugated side can be
back-scattered in such a way that it interferes constructively
with the one propagating in the opposite direction, thus obtaining
a localized unidirectional SPP source. Here, we provide a
comprehensive version of our proposal and discuss in some extent
its range of validity. Additional experimental measurements will
be also presented.

This paper is organized as follows: in \Sref{descrip} we summarize
the key concepts of our proposal and focus in some quantitative
aspects of SPP generation and reflection. The validity of our
simple wave interference model is discussed in \Sref{valid}.
Finally, experimental results are presented in \Sref{exp}, prior
to general conclusions.

\section{Description of our proposal}
\label{descrip}

\begin{figure}
  \flushright
  \includegraphics[width=0.84\textwidth]{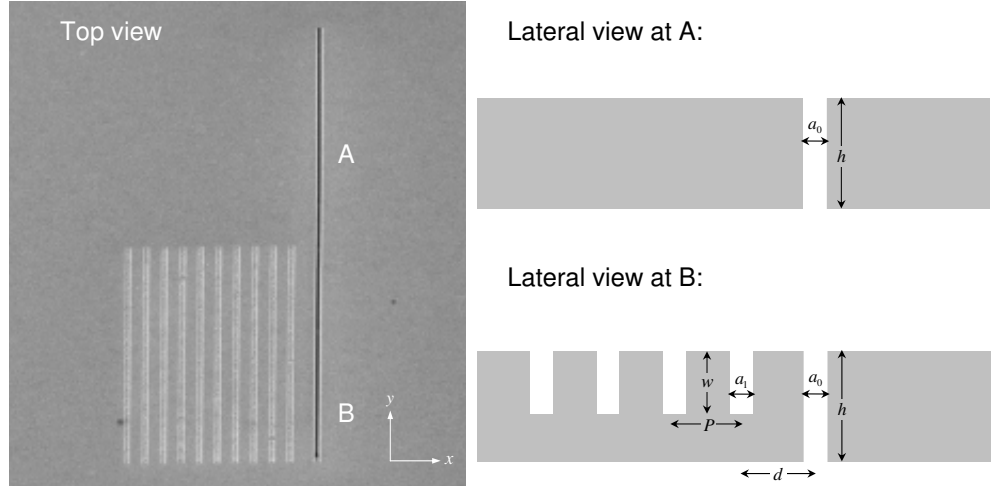}\\
  \caption{Scanning electron micrograph and schematic diagrams of the proposed structure. Parameters
  $\{a_i,h,w,d,P\}$ defining the geometry of the system are also shown.}\label{fig1}
\end{figure}

A picture of the proposed structure is shown in \fref{fig1}. A
periodic array of one-dimensional (1D) indentations is fabricated
at the output metal surface close and parallel to the illuminated
slit. The starting point for such a design can be found in a
previous work on 1D SPP scattering by means of a modal expansion
formalism \cite{flt05,flt07apa}. In order to cope with SPP
launching, we considered a single slit flanked by an array of
indentations (rectangular grooves) placed in the output surface of
a thick metallic film. Eventually, the distance between the slit
and indentations was taken to be infinity. In this way, the slit
merely played the role of a theorist's SPP-launcher, as far as it
can be shown that the field created by the slit corresponds to SPP
illumination into the grooves. Besides, we also found a simple
geometrical condition for the groove array to behave as a perfect
Bragg mirror, associated to the low-$\lambda$ edge of the
plasmonic band gap for the periodic system. Combining these two
elements, one can obtain a remarkably simple scheme to modulate
the SPP coupling-in at a real back-side illumination experiment:
given an incident wavelength, let us design a groove array for
which SPP reflectance rises to a maximum and place it at a
distance $d$ from the slit (Situation B of \fref{fig1}). Hence,
any outgoing SPP generated at the same side of the slit will be
mainly back-scattered by the grooves and interfere either
constructively or destructively with the one that is generated at
the opposite side. This interference can be tuned by adjusting the
separation $d$ between the slit and the first groove of the array,
defined centre to centre. The total phase difference, $\phi$,
between the interfering SPPs will then consist of the phase change
upon reflection plus the additional shift resulting from the two
different path lengths along the metal:
\begin{equation}
\phi=\phi_R +2 Re[k_p] d,  \label{eqphi}
\end{equation}
where $k_p$ holds for in-plane plasmon wave-vector. According to
\eref{eqphi}, constructive or destructive interference should
occur for those phase values equal to, respectively, even or odd
multiples of $\pi$.

It is clear that, as  will be discussed in \sref{valid}, several
objections may arise against this very simplified model, but
before we turn to its validity, let us take a closer look at the
two ingredients on which it is based: the generation of SPPs at a
sub-wavelength aperture and the phase they acquire as a result of
Bragg reflection.

\subsection{SPP generation at a single sub-wavelength
slit}\label{sppgen}

\begin{figure}
  \flushright
  \includegraphics[width=0.84\textwidth]{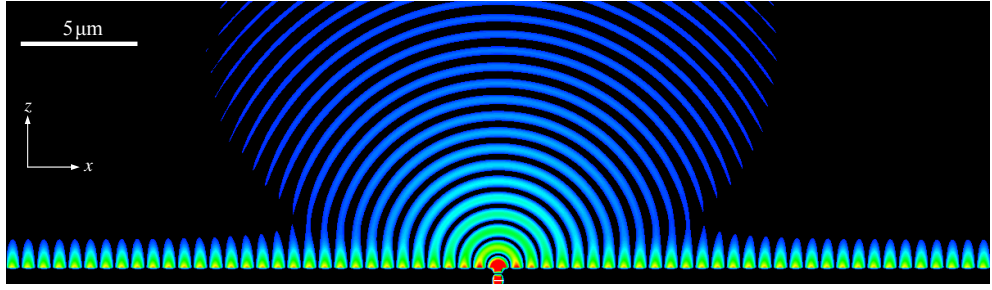}\\
  \caption{Calculated $|Re[H_y]|$ distribution at the output surface of
  an Au film perforated by a single slit (Situation A of \fref{fig1}). Incident light is $p$-polarized and
  impinges normally onto the back side of the metal surface at $\lambda = 800 $ nm . Here, slit width $a_0= 160$ nm
  and film thickness $h=300$ nm.}\label{fig2}
\end{figure}

\Fref{fig2} renders a  finite-difference time-domain (FDTD)
\cite{Taflove2000} simulation of the electromagnetic (EM) field
distribution originated from $p$-polarized back-side illumination
of a sub-wavelength slit on a thick Au film. As can be seen, the
most of the field is diffracted away, but a significant fraction
appears to be bound to the metallic surface at each side of the
aperture. However, to what extent is that a confined state of
surface plasmon?

Such an assessment requires that analytical expressions for the EM
field distribution created by the slit be obtained without any a
priori assumption about the presence of SPPs. For that purpose, we
have made use of the above-mentioned modal expansion technique.
Given that it has been extensively described elsewhere
\cite{flt05,flt07apa}, let us just briefly summarize its basic
ingredients: the EM fields are expanded in terms of the eigenmodes
in each spatial region (plane waves at input/output regions and
waveguide modes inside the indentation) and then the expansion
coefficients are obtained by just matching appropriately the
parallel components of the fields at the two metal-dielectric
interfaces. The dielectric response of the metal is taken into
account by applying surface impedance boundary conditions (SIBC)
\cite{Jack_book} to the tangential components of the EM fields at
the metallic surface. For a non-magnetic medium,
\begin{equation}\label{contineq}
\bi{F}_{t}(\bi{r}) \equiv \bi{E}_{t}(\bi{r}) -Z_s
\bi{H}_{t}(\bi{r})\times\bi{n}(\bi{r})=0 ,
\end{equation}
where $Z_s=\varepsilon(\lambda)^{-1/2}$ and $\bi{n}(\bi{r})$ is
the unitary vector normal to the surface directed into the metal
half-space. However, SIBC are not applied at the vertical walls
defining the slit but for the calculation of propagating constants
along the $z$ direction. This choice allows us to express the EM
fields inside in terms of the waveguide eigenmodes of a perfect
conductor (PC), which are known analytically. Although the
absorption inside the cavities is therefore neglected, one can
expect this not to be a serious shortcoming when considering sizes
much greater than the skin depth. The end product of our expanding
and matching is a linear system of algebraic equations that
connect the modal amplitudes of the $\bi{F}$ field at the input
and output openings of the slit. Once those self-consistent
amplitudes are found, it is straightforward to obtain the EM
fields at any desired point.

By imposing the constraint that incident light impinge in
``classical mounting'' (ie. within the $xz$ plane), we just have
to concern ourselves with the $y$-component of the magnetic field.
From a mathematical point of view, $H_y$ at the output side is
obtained by integrating all across the slit every considered
eigenmode $\phi_n$ multiplied by a scalar 1D Green's function and
then weighting each contribution with the corresponding amplitude
$E'_n$ at the output opening:
\begin{equation}
H_y(x,z)=- \sum_n E'_n
\int^{x_0+a_0/2}_{x_0-a_0/2}dx'G(x,x';z)\phi_n(x'),\label{def_hy}
\end{equation}
where $z$ stands for the distance from the output surface. This
closely resembles the Huygens-Fresnel description of wave
propagation in terms of a set of punctual emitters, but we have to
keep in mind that all those ``emitters'' are self-consistently
connected.

However, the information on the character of the generated field
is contained neither in the modes nor in their amplitudes, but in
the propagator itself:
\begin{equation}
G(x,x';z) = \frac{i}{\lambda}\int^{+\infty}_{-\infty}dk
\frac{\exp[i(k (x-x')+\sqrt{k_0^2-k^2}z)]}{\sqrt{k_0^2-k^2} +
k_0Z_s}, \label{def_g}
\end{equation}
where $k_0 \equiv 2 \pi/ \lambda$.
Despite its impressive appearance, $G(x,x';z)$ just computes the
projection of EM fields at the opening of the slit onto all
possible diffracted waves, whether they are propagating or
evanescent. Bound-to-interface contributions are incorporated into
the picture as a consequence of finite $Z_s$, which makes the
difference with respect to perfect conductor approximation: for
$Z_s = 0$, \eref{def_g} transforms into an integral representation
of the 0th-order Hankel function of the first kind (ie. the
well-known Green's function for two-dimensional Helmholtz operator
\cite{Arfken_book}), otherwise it has to be evaluated numerically.
Such numerical inspection reveals that $G(x,x';z)$ tends to the PC
result for $|x-x'|<<\lambda$ irrespective of $Z_s$
\cite{flt05,flt07apa}. On the other hand, in the regime where $
z,|x-x'| \approx O(\lambda) $, oscillatory contributions within
the kernel of \eref{def_g} mutually cancel everywhere but in the
region close to the integrand singularities at $k=\pm k_p$, with
$k_p$ satisfying
\begin{equation}
\sqrt{k_0^2-{k_p}^2}=-Z_s k_0.
\end{equation}
This is, by the way, the SPP dispersion relation of a flat
metal-dielectric interface within the SIBC. In that asymptotic
limit, the Green's function can be explicitly approximated as
\begin{equation}
G_{as}(x,x';z) = -\frac{k_0^2 Z_s}{k_p}\,e^{i(k_p|x-x'|-k_0 Z_s
z)},\label{gasint}
\end{equation}
Therefore, and even in the presence of absorption, SPPs govern the
EM coupling along the surface at a distance of several
wavelengths, whereas ``PC-like'' behaviour is observed at the
close vicinity of the slit. It is worth mentioning that this
simple fact is completely misinterpreted in several recent papers,
as pointed out in a previous work \cite{fjcommentnp06}. In any
case, the existence of these two regimes have also been remarked
by introducing a ``creeping wave contribution'' that rapidly
vanishes for increasing distances and is explicitly defined as the
difference between total and SPP fields along the metal-dielectric
interface \cite{Lalanne06np}.

\begin{figure}
  \flushright
  \includegraphics[width=0.84\textwidth,angle=-90]{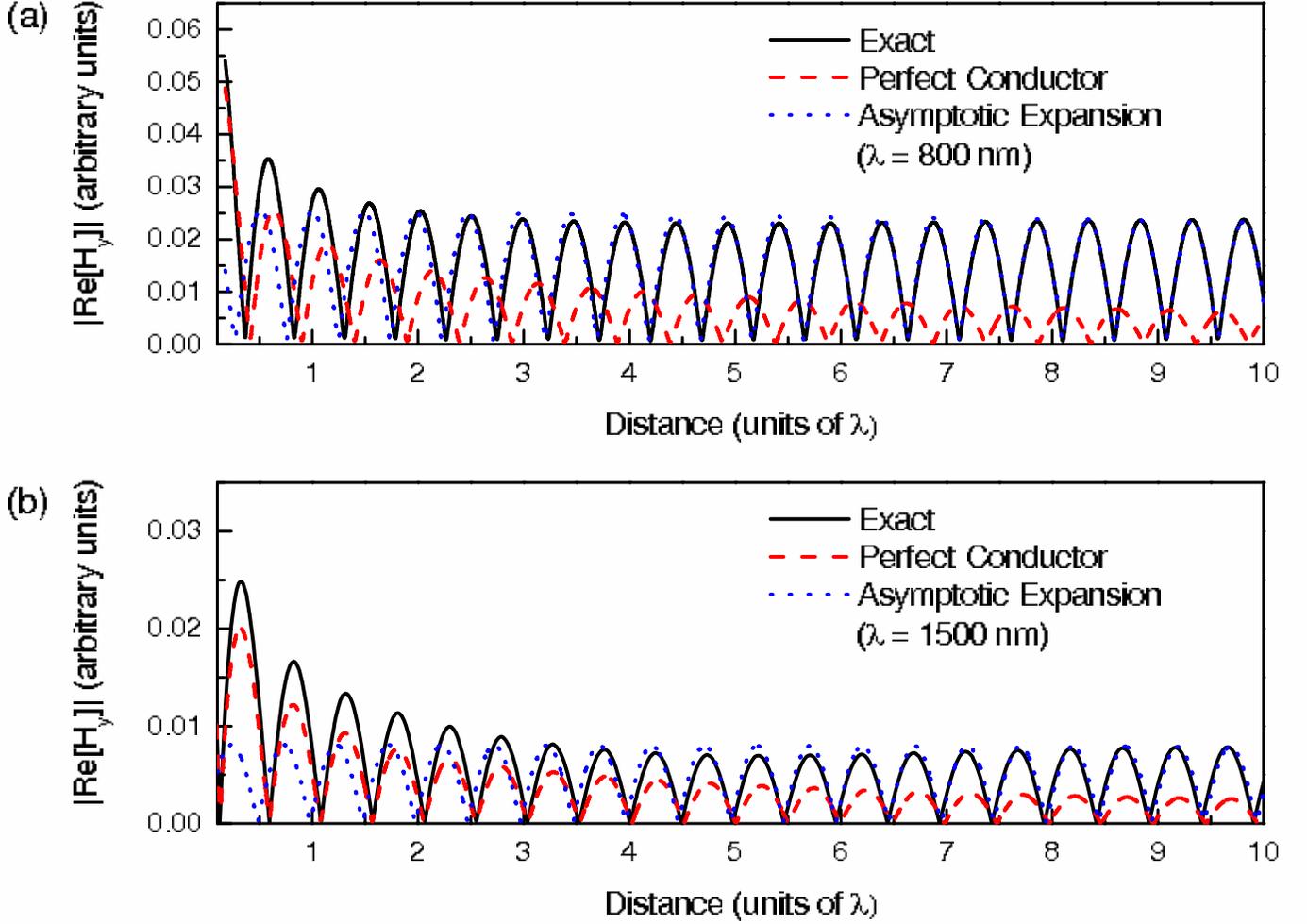}\\
  \caption{Calculated $|Re[H_y]|$ as a function of the distance from
  the center of the slit, evaluated at the output surface of an Au film.
  Solid lines represent the full calculation, whereas dashed and dotted
  ones stand for perfect conductor approximation and asymptotic expansion,
  respectively. The geometrical parameters are: slit width $a_0= 160$ nm
  and film thickness $h=300$ nm. Incident light is $p$-polarized and
  impinges normally onto the back-side of the metal surface. Panels (a) and (b) show
  results for $\lambda=800$ nm and $\lambda=1500$ nm, respectively.}\label{fig3}
\end{figure}

In order to determine the precise range of distances for the EM
field at \fref{fig2} to be dominated by either ``PC-like'' or SPP
contribution, we have calculated $|Re[H_y]|$ at the metal surface
for $Z_s$ values corresponding to that of Au at 800 and 1500 nm.
Each calculation was carried out for the exact, asymptotic and PC
versions of Green's function. As can be seen in \fref{fig3},
comparison with the exact result in the near-infrared (NIR) shows
that the asymptotic limit is already reached for a distance of
about $2\lambda$ from the center of the slit, which is increased
up to $6\lambda$ when the incident wavelength falls within the
telecom range. Consequently, it is only for greater distances that
we can unambiguously establish a one-to-one correspondence between
fields at the interface and SPPs.

\begin{figure}
  \flushright
  \includegraphics[width=0.84\textwidth]{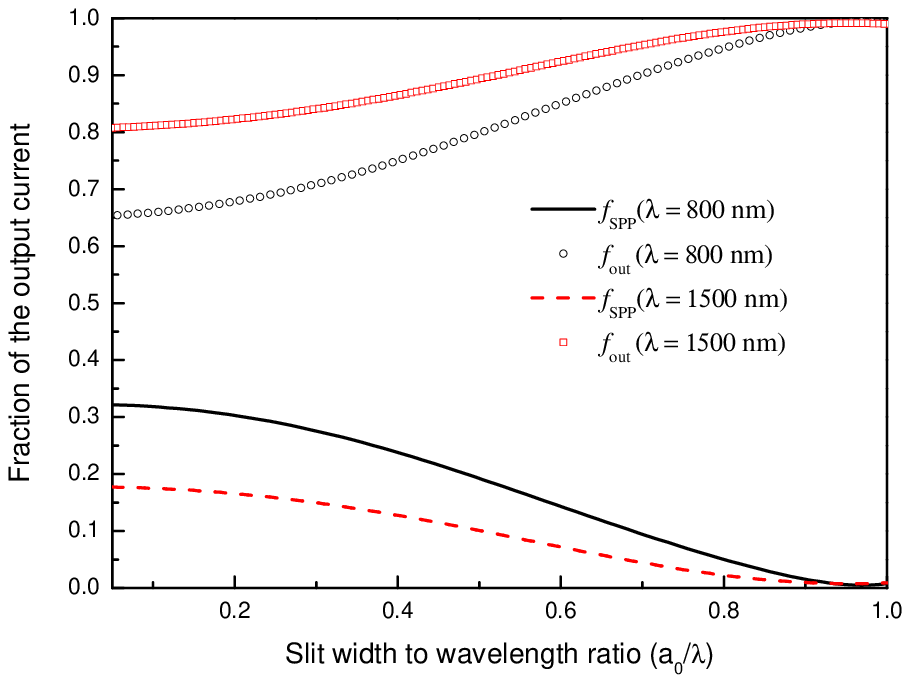}\\
  \caption{Fraction of the energy that is transferred into
  SPPs (lines) and scattered out of the plane (symbols)
  at the output surface of an Au film perforated by a single slit
  that is back-side illuminated with p-polarized light. The values for $f_\mathrm{SPP}$
  are calculated at distances of $\pm a_0/2$
  from the center of the slit.}\label{fig4}
\end{figure}

In \fref{fig4} we present the fraction of the output current that
is transferred into SPPs ($f_\mathrm{SPP}$) and scattered out of
the plane ($f_\mathrm{out}$) for the same $Z_s$ parameters as in
\fref{fig3} all across the sub-wavelength regime.
Given that SPPs gradually attenuate when propagating along the
metal, the values for $f_\mathrm{SPP}$ are calculated at $x=\pm
a_0/2$ in order to compare with those of $f_\mathrm{out}$. As the
slit width increases, the out-of-plane radiation is clearly
favored at the expense of the coupling into SPPs, which can be
easily found to be proportional to $(\sin [k_p\, a_0/2]/k_p\,
a_0)^2$ because of the geometry of the system
\cite{flt05,flt07apa}.
For typical experimental width $a_0=160$ nm, no more than $30\%$
of the output energy is driven into SPPs at $\lambda=800$ nm and
such percentage is reduced to $17\%$ at $\lambda=1500$ nm. These
values are in good agreement with those previously reported
\cite{Lalanne2005} and provide a preliminary estimate of the
expected performance for our proposed slit+grating structure when
operating at perfect constructive interference condition.

\subsection{Phase shift upon Bragg reflection}
\label{phaseshift}

\begin{figure}
  \flushright
  \includegraphics[width=0.84\textwidth]{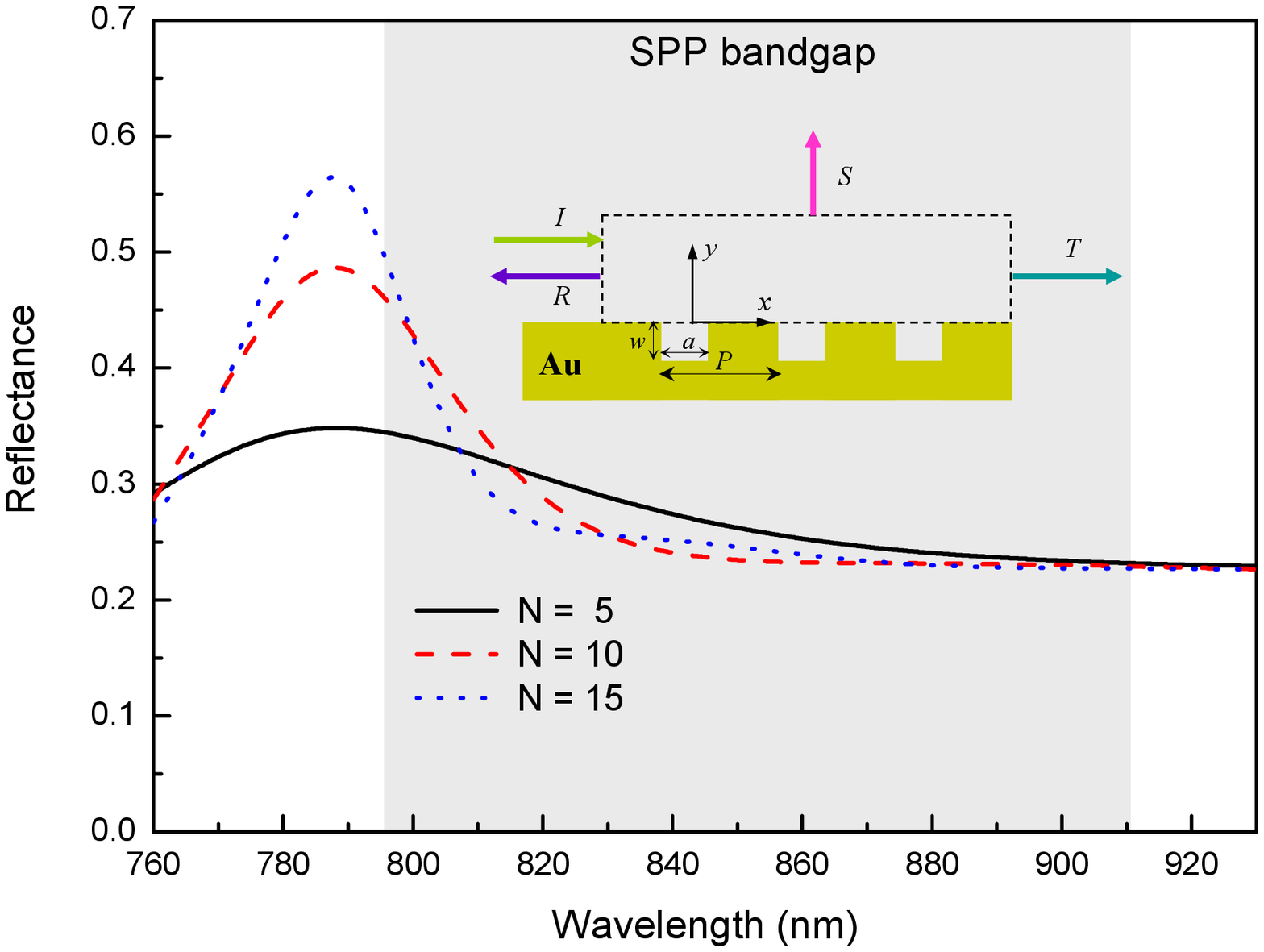}\\
  \caption{Calculated reflectance of SPPs by a finite periodic array consisting of
  5 (solid line), 10 (dashed) and 15 (dotted) grooves carved on Au. Here, $a=w=100$ nm
  and $P=390$ nm. Gray-shaded area marks the region where a plasmonic bandgap occurs.
  SPP fields are evaluated at $x=-3.5 \mu$m $(\approx -3.8\lambda_{max}$), the origin being located
  at the center of the first groove.}\label{fig5}
\end{figure}

As mentioned at the beginning of the section, it has been shown
that the reflection of SPPs by a periodic array of indentations
presents maxima at those frequencies corresponding to the
low-$\lambda$ edges of plasmonic bandgaps \cite{flt05,flt07apa}.
For narrow sub-wavelength indentations, the spectral locations of
these edges can be approximated by folding the dispersion relation
of SPPs for a flat metal surface into the first Brillouin zone
\cite{Kitson96}. Within the SIBC, such folding results in
\begin{equation}\label{lambdaRmax}
k_pP=k_0 Re[q_p] P = m \pi, \quad m=1,2 \ldots
\end{equation}
where $P$ is the period of array and  $q_p \equiv \sqrt{1-Z_s^2}$.
Remarkably, although the reflectance maxima depend on the groove
geometry (width and depth) and the number of grooves
\cite{flt05,flt07apa}, their spectral locations do not (see
\fref{fig5}).

\begin{figure}
  \flushright
  \includegraphics[width=0.84\textwidth]{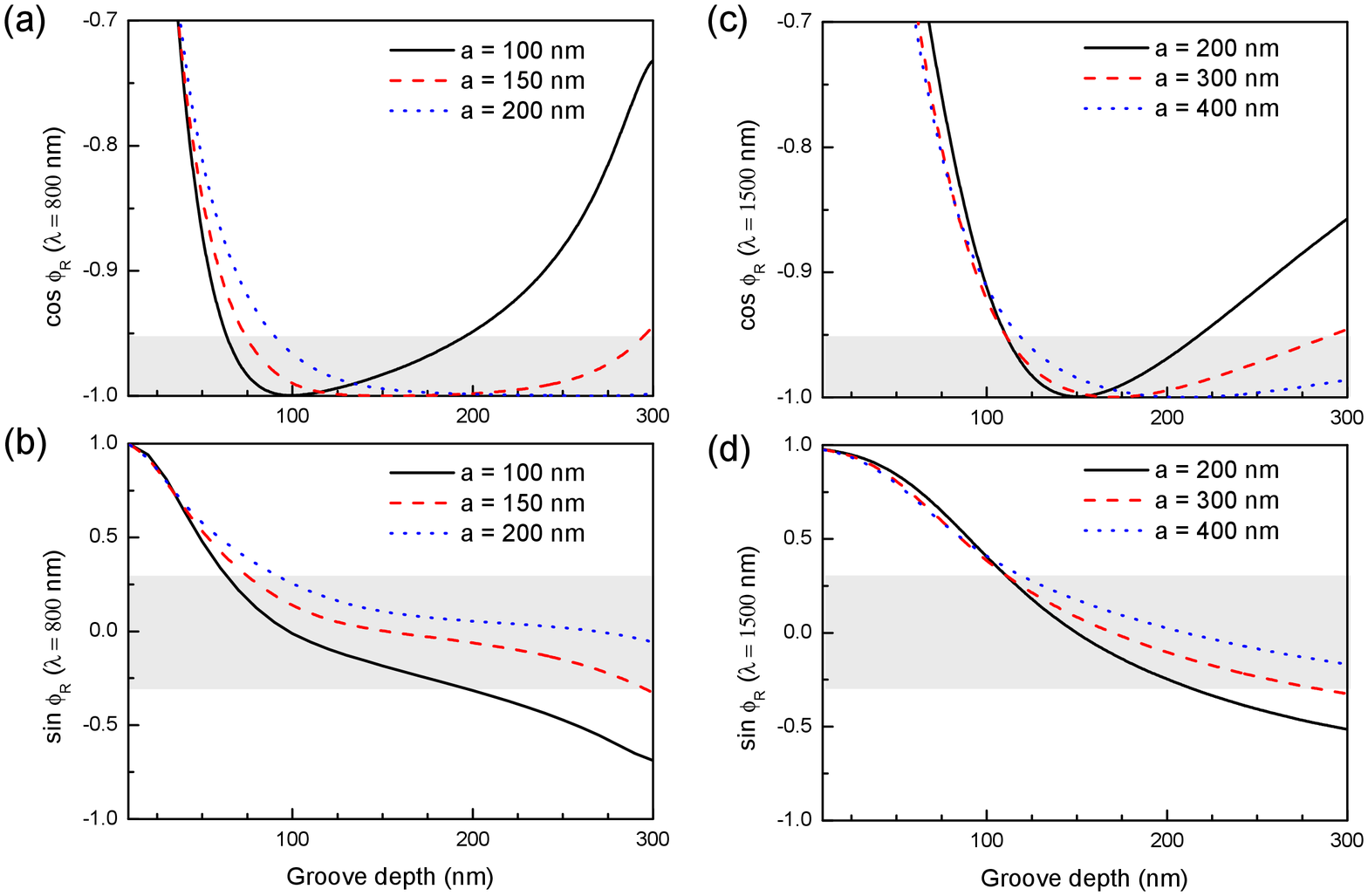}\\
  \caption{Calculated values of $\cos \phi_R(\lambda_R), \sin \phi_R(\lambda_R)$
  of a 10-groove array in Au for increasing values of groove depth.
  Panels (a) and (b) display results for $a=100, 150$ and 200 nm
  (solid, dashed and dotted lines, respectively) evaluated at $\lambda_R=800$ nm
  ($P=390$ nm, m = 1). Results for $a=200, 300$ and 400 nm at
  $\lambda_R=1500$ nm ($P=750$ nm, m = 1) are presented in panels (c) and (d). Gray-shaded
  areas mark the region where $\phi_R=\pi \pm 0.1\pi$. Results for $\lambda_R=800$ nm
  and $\lambda_R=1500$ nm are calculated at distances of $3\lambda$ and $7\lambda$
  from the center of the first groove, respectively.}\label{fig6}
\end{figure}

Assuming that $\lambda$ and $P$ fulfill \eref{lambdaRmax}, let us
consider the phase shift for a given resonant wavelength
$\lambda_R$. Information on such shift is contained in the complex
reflection coefficient $r$ relating the amplitudes of incident and
reflected fields. Although the obtention of $r$ is usually
regarded as a mere preliminary to that of reflectance (defined as
$R=|r|^2$), we can always establish a straightforward connection
between $r$ and phase shift $\phi_R$:
\begin{equation}
\cos \phi_R = Re[r]/|r|;\quad \sin \phi_R = Im[r]/|r|.
\end{equation}
Once the asymptotic limit is already reached, these auxiliary
magnitudes  $\cos \phi_R, \sin \phi_R$ provide complete
information about SPP shift upon reflection, irrespective of the
exact distance at which fields are evaluated. We have found that
$\phi_R$ is close to $\pi$ over a wide range of groove depths for
$a / \lambda \leq 0.2$ at both NIR and telecom ranges, as can be
seen in \fref{fig6}. Taking this result into account and
substituting for $k_p$ from \eref{lambdaRmax} into \eref{eqphi}
yields
\begin{equation}
\phi(\lambda_R)=(2md/P+1)\pi \label{eqphi_R},
\end{equation}
which reduces the design of our proposed scheme to a suitable
choice of the $d/P$ ratio.

\section{Validity of the simple wave interference model}
\label{valid}

Our previous discussion leading to \eref{eqphi_R} implies that
slit and grating be considered as independent elements. Therefore,
it does not take into account the radiation coming back from the
grooves, while, in principle, EM-fields at all openings have to be
self-consistently calculated \cite{lmm03}. In order to quantify
the ``perturbation'' of the SPP source (ie. the slit), we define a
re-illumination parameter $\xi$ that averages the modification of
the $x$-component of the electric field inside the slit originated
by the adjacent grating:
\begin{equation}
\xi = \frac{1}{a_0} \int^{+a_0/2}_{-a_0/2} dx' \vert
1-E_x(x')/E_x^{ss}(x')\vert,\label{xidef}
\end{equation}
where $a_0$ is the width of the slit, $E_x$ the $x$-component of
the electric field calculated in the presence of the array and
$E_x^{ss}$ the one obtained for the isolated slit.

\begin{figure}
  \flushright
  \includegraphics[width=0.84\textwidth]{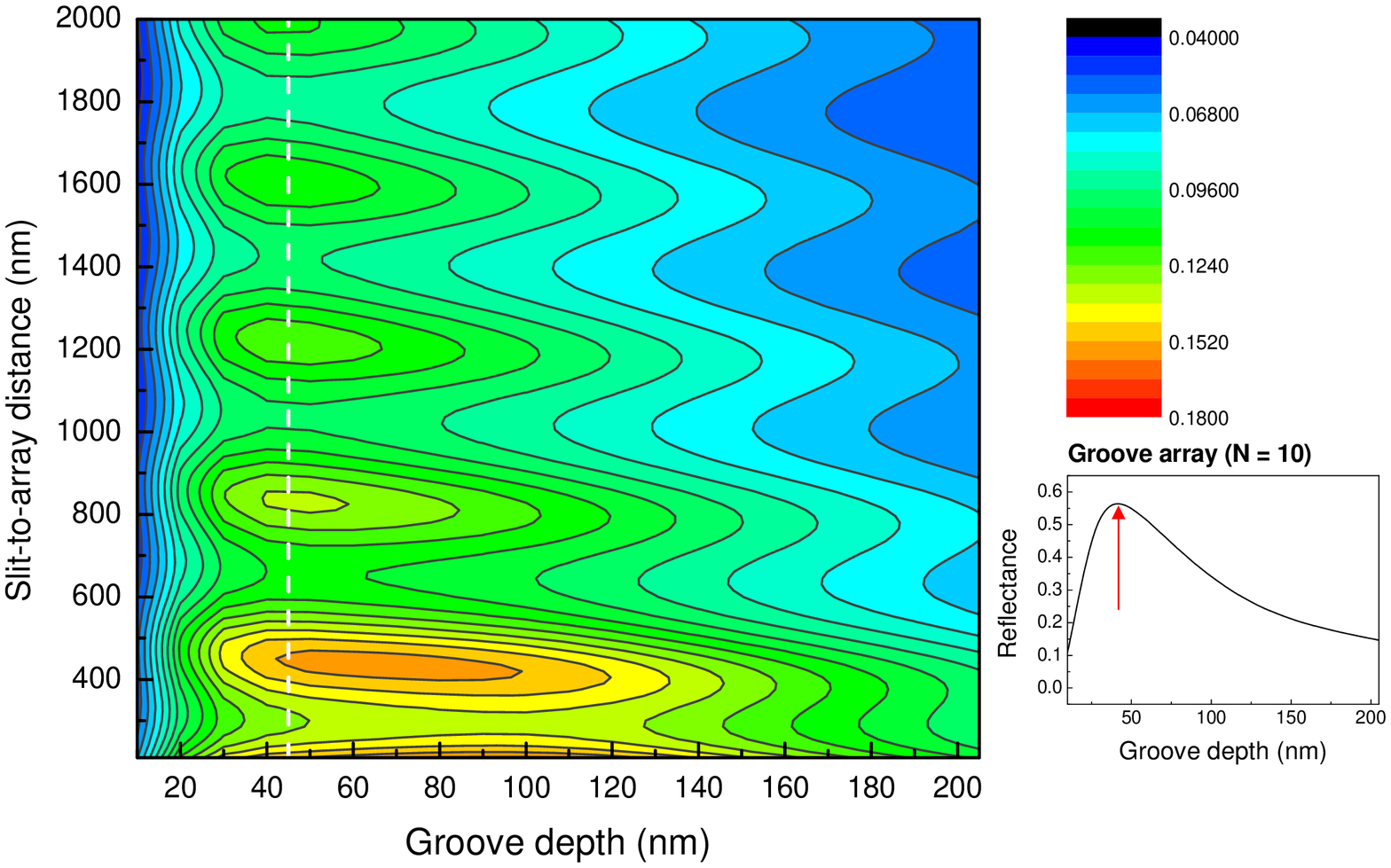}\\
  \caption{Contour plot of re-illumination parameter $\xi$ in Au as a function of groove depth
  and slit-to-array separation. Here $a_0=a_1=160$ nm, $P=390$ nm, $N =10$ and $\lambda=800$
  nm. Vertical dashed line marks the maximum value of the reflectance curve for the isolated groove array,
  which is presented at the lower right inset panel.}\label{fig7}
\end{figure}

In \fref{fig7} we present a contour plot of $\xi$ vs groove depth
and slit-to-array separation for a system with 10 grooves at
$\lambda=800$ nm. As can be seen, the modification of the field
pattern within $[400,\,800]$ nm is below $15\%$ and $\xi$ rapidly
decreases for increasing distances, thus supporting our implicit
assumption in \eref{eqphi_R}. With respect to the dependence on
groove depth, it is governed by the reflectance properties of the
array, $\xi$ rising to its maximum as $R$ does (see inset in
\fref{fig7}). Such a maximum becomes clearer the more separation
approaches to the plasmonic regime ($d \approx 3\lambda$). On the
other hand, modulation along the vertical axis results from simple
interference between counter-propagating SPP waves originated at
the slit and its nearest groove. Therefore, sequential minima of
$\xi$ appear for $d=(2m+1)\lambda_p/4$, whereas $\{\xi_{max}\}$
are associated with $d=m\lambda_p/2$, given that
$\lambda_p=2\pi/k_p$ and $m=0,1,2 \ldots$

However, the key point of our proposal still relies on SPPs being
reflected by a groove array, while the EM fields radiated by the
slit cannot be considered ``purely plasmonic'' but at a distance
of several wavelengths (see \fref{fig3}). In order to characterize
the efficiency  of the sli+array system as an SPP-launcher for any
slit-to-array separation, we introduce its ``efficiency ratio",
$E_R$: given that the array be located at the left side of the
slit (see \fref{fig1}), $E_R$ is defined as the quotient between
the current intensity of right-propagating SPP with and without
the grooves. Strictly speaking, $E_R$ provides the efficiency of
the output side of the device. The total efficiency, defined as
the percentage of incident energy transferred onto the plasmon
channel, strongly depends on the illuminating setup. $E_R$ should
vary within the interval $[0,4]$ showing a dependence on the
distance between the illuminating slit and the groove array. More
importantly, $E_R>2$ implies that the right-propagating SPP
current in the presence of grooves is larger than the total SPP
current (left- plus right- moving) in the single slit case, so
some of the power radiated out of plane is redirected onto the SPP
channel. According to our simple wave interference model,
\begin{equation}
E_R \approx |1+re^{2ik_pd}|^2,\label{ERmodel}
\end{equation}
where $r$ is the complex reflection coefficient of the groove
array for SPPs.

\begin{figure}
  \flushright
  \includegraphics[width=0.84\textwidth]{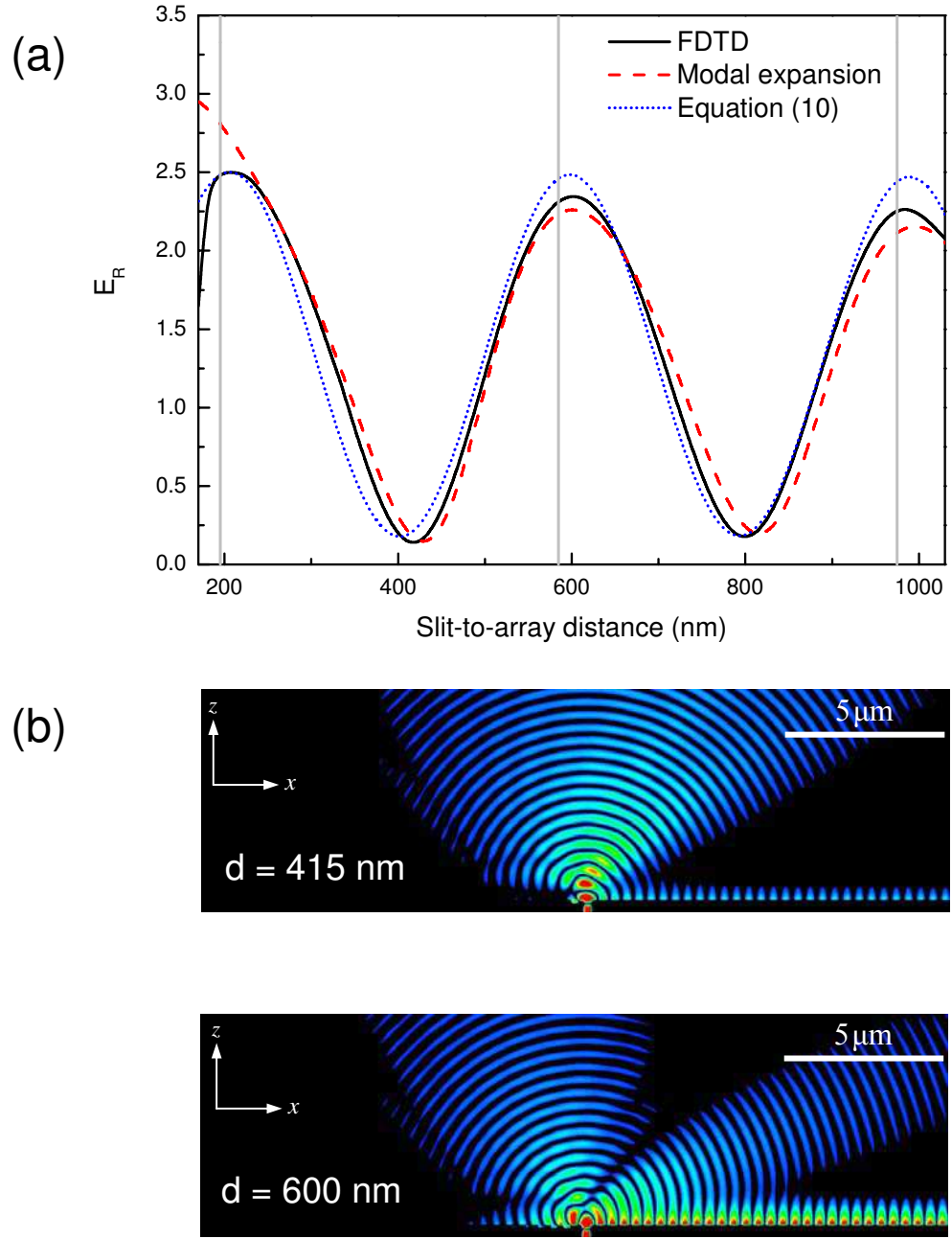}\\
  \caption{Numerical results for the SPP launcher at
  wavelength $\lambda=800$ nm. \textbf{(a)} Dependence of the efficiency ratio
  $E_R$ on the slit-to-array distance. The geometrical
  parameters defining the system are: slit and groove widths $a=160$ nm,
  groove depth $w=100$ nm and array period $P=390$nm. The figure renders the curves obtained
  by means of FDTD (solid), modal expansion (dashed) and equation \eref{ERmodel} (short-dotted).
  Vertical lines mark the positions of $E_R$ maxima according to \eref{eqphi_R}.
  \textbf{(b)} Calculated $|Re[H_y]|$ distributions over $xz$ plane for two different distances
  corresponding to minimum and maximum values of $E_R$ at $\lambda=800$ nm.}\label{fig8}
\end{figure}

To check the validity of \eref{eqphi_R} and \eref{ERmodel} for
slit-to-array separations outside the asymptotic regime, we have
carried out numerical calculations of EM fields by means of both
modal expansion and FDTD. The system under consideration is
intended to operate at a wavelength of 800 nm on a gold film
\cite{Vial2005}. We consider an array of $10$ grooves with a
period $P=390$ nm. The depth of the grooves is chosen to be
$w=100$ nm, while the width of both grooves and slit is $a=160$
nm, which are typical experimental parameters. \Fref{fig8}(a)
shows the comparison between \eref{ERmodel} and numerical
evaluations of $E_R$, as well as the location of interference
maxima (vertical lines) predicted by \eref{eqphi_R} for $m=1$. The
agreement between the modal expansion and FDTD results is
excellent but for distances at which intra-wall coupling between
the slit and the first groove has to be taken into account ($d
\approx 2a$). As can be seen, the locations of maximum $E_R$ are
accurately predicted by \eref{eqphi_R}, which allows us to design
SPP-launchers without elaborate numerical calculations. Moreover,
the simplified model of \eref{ERmodel} provides a good
approximation to $E_R$ with the sole input of $r$. This also
implies that non-plasmonic contributions to groove illumination
play a minor role in the occurrence of either constructive or
destructive interference, which is clearly described by
\eref{ERmodel} with the except of minor shifts.

In addition to the efficiency ratio, field patterns in both
minimum and maximum condition were also calculated using the FDTD
method. As shown on \fref{fig8}(b), SPPs are completely absent
from the left side of the slit whereas field intensity at its
right side is clearly modulated by the slit-to-array separation,
which also governs the spatial distribution of the field that is
radiated into the vacuum.

\section{Experimental results}
\label{exp}

\subsection{Near-infrared measurements}\label{NIR}

For our proposal to be tested out at the NIR regime, several
slit+array samples were fabricated on gold films with a Focused
Ion Beam (FIB). As described in \cite{flt07np}, each sample
consists of a single long ($L=30 \mu$m) slit of width $a_0=160$ nm
perforated at a 300-nm-thick film that is flanked by a periodic
array of grooves ($P=390$ nm, $a=160$ nm, $w=100$ nm). Such array
is placed at a given distance $d$ and only extends over $L/2$ (see
\fref{fig1}). This kind of samples enables us to measure $E_R$, as
the upper part can be used as an in-chip reference of the
``isolated slit''.

\begin{figure}
  \flushright
  \includegraphics[width=0.84\textwidth]{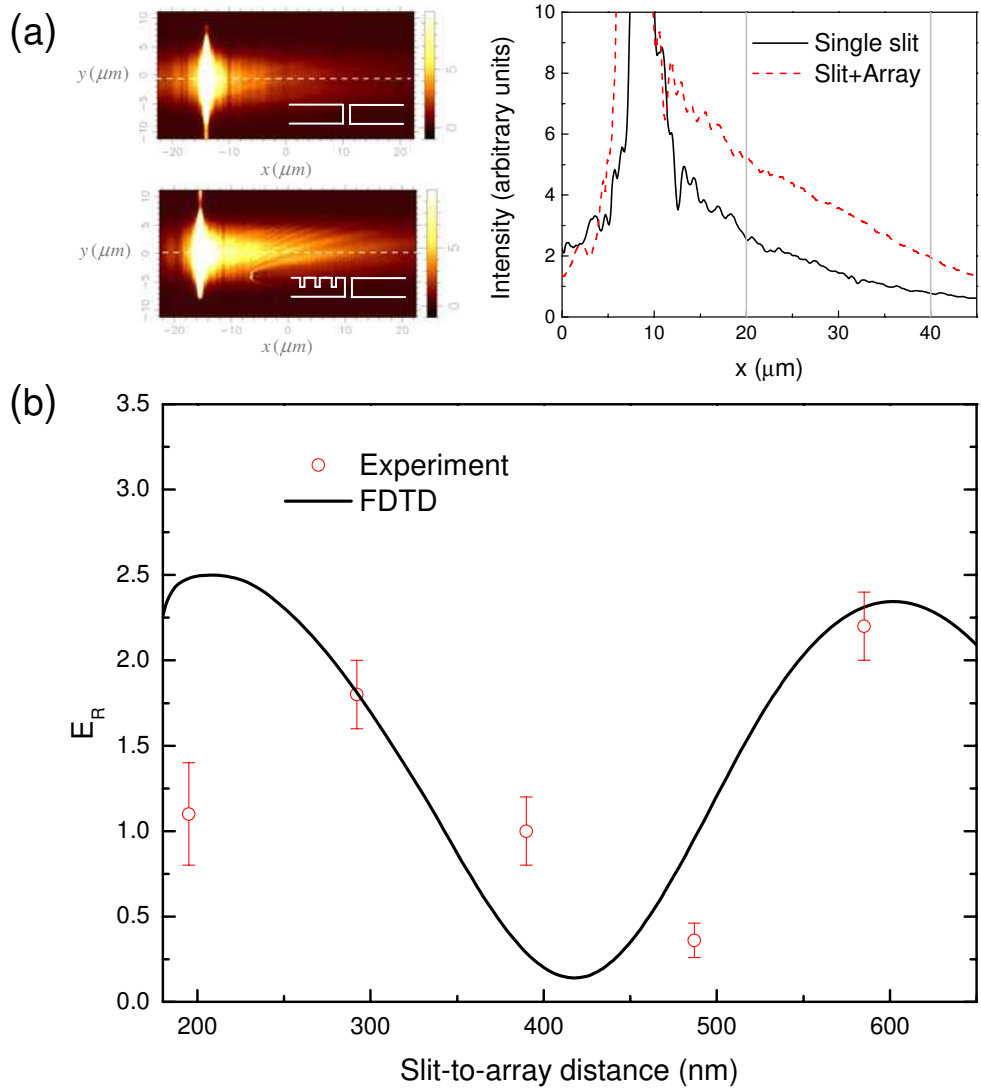}\\
  \caption{Experimental measurement of $E_R$ at $\lambda=800$ nm for the same
  geometrical parameters as in \fref{fig8}. \textbf{(a)} PSTM micrographs recorded for a sample with $d=585$ nm at
  both ``single slit'' (top) and slit+array configurations (bottom).
  The right panel shows the two cross-cuts from which $E_R$ is obtained.
  Vertical lines define the interval along the ratio is
  averaged. \textbf{(b)} Experimental (circles) and numerical (solid line) values of $E_R$
  as a function of slit-to-array distance. The error bars represent the standard deviation over a set of
  different structures with the same nominal parameters.}\label{fig9}
\end{figure}

A set of samples with $d=\{195,292,390,486,585\}$ nm was imaged at
800 nm by a Photon Scanning Tunneling Microscope (PSTM) making use
of an incident focused beam illumination. For each sample, a pair
of images was recorded by scanning at a constant distance of about
$60$ to $80$ nm from the surface (see \fref{fig9}(a)). The first
image of the pair, corresponding to a SPP generated from a single
slit, is obtained by focusing the laser beam on the upper part of
the slit. For the second image, the laser beam is moved to the
lower part in order to collect the data for the slit+array
structure. An average longitudinal cross-cut of each image is
obtained by using $20$ longitudinal cross-cuts, corresponding to
different coordinates along the slit axis. Then, the relative
position of the two average cross-cuts is adjusted so that the
saturated areas (i.e. the signal taken right on top of the slit)
are super-imposed. Finally, the experimental efficiency ratio,
$E_R$, is extracted by averaging the ratio between the two curves
along the longitudinal cross-cut. \Fref{fig9}b renders
experimental values (circles) of $E_R$ for the five different
samples fabricated, as well as the ones obtained from FDTD
simulations (solid line). The concordance between measurements and
theoretical predictions is quite remarkable, especially when
taking into account that each experimental point corresponds to an
average over a different set of samples. We find that this
agreement (previously reported in \cite{flt07np}) provides a clear
support to our proposal for a localized unidirectional SPP source.

\begin{figure}
  \flushright
  \includegraphics[width=0.84\textwidth]{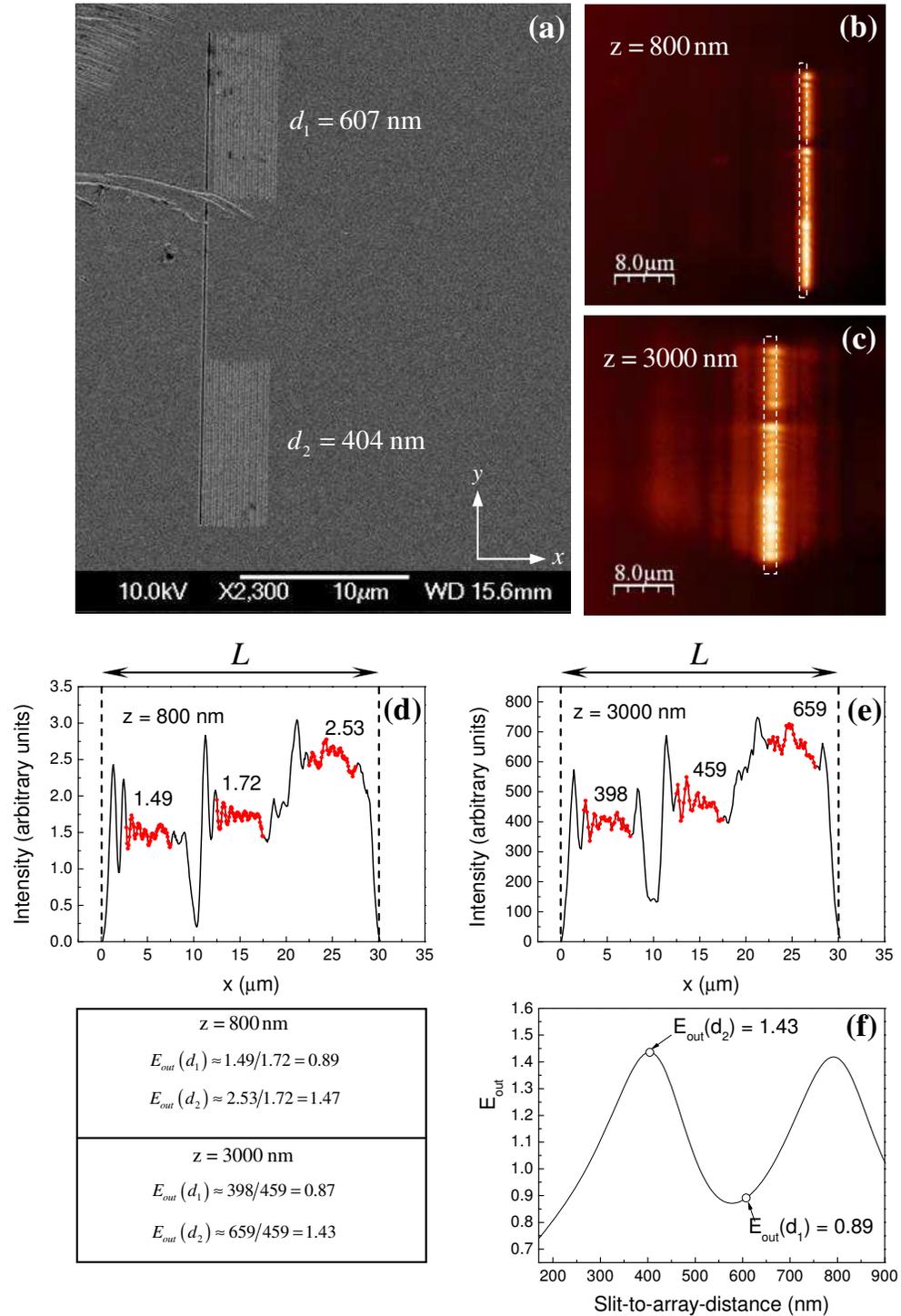}\\
  \caption{Experimental estimate of $E_{out}$ at $\lambda=800$ nm.
  \textbf{(a)} Scanning electron micrograph of the sample. The geometrical
  parameters are: slit length $L=30 \mu$m, slit width $a_0= 104$ nm,
  groove width $a=75$ nm, groove depth $w=100$ nm and array period $P=390$
  nm. \textbf{(b), (c)} PSTM micrograph recorded at distances of  800, 3000 nm from the metal surface.
  \textbf{(d), (e)} Average longitudinal cross-cuts along dashed rectangles in images (b) and
  (c). Vertical dashed lines mark the position of the slit. \textbf{(f)} Calculated $E_{out}$ as a function of slit-to-array distance. Lower left table:
  summary of the experimental estimates of $E_{out}$.}\label{fig10}
\end{figure}

Another way of looking at the role of surface corrugation is to
consider its influence on the fraction of the output energy that
is radiated into vacuum. Given that some of the radiated power is
redirected onto the SPP channel for the condition of maximum $E_R$
(see field pattern at \fref{fig8}), we may wonder whether or not
the radiated field is also modulated by the slit-to-array
separation. For that purpose, a new magnitude $E_{out}$ can be
defined as the ratio between the radiated energy with and without
the grooves. According to our numerical simulations, such
``out-of-plane efficiency'' presents a similar (but opposite)
dependence on $d$ to that of $E_R$. In order to obtain
experimental values for $E_{out}$, a new type of sample was
designed (see \fref{fig10}(a)). Now, the illuminating slit is
flanked by two groove arrays with the same periodicity $P=390$ nm,
each one extending over $L/3$. No corrugation is present at the
middle part of the system, for it to be used as the ``single
slit'' reference. The array on the top is located at a distance
$d_1=607$ nm for which the coupling to SPPs rises to a maximum at
$\lambda=800$ nm, whereas a minimum appears for the distance
$d_2=404$ nm of the bottom one. Consequently, the far-field
radiation pattern of the composed structure is expected to present
a $d_1 \rightarrow d_2$ ascending staircase profile.

In \fref{fig10}(b), (c) we present PSTM images recorded at 800 and
3000 nm from the surface of the sample. As can be seen, the
intensity distribution along the illuminating slit increases from
the upper to the middle third, as well as from the middle to the
lower. Although this behavior is in qualitative agreement with our
predictions, a rigorous determination of $E_{out}$ would have
required extensive measurements similar to those of $E_R$.
Unfortunately, such procedure became impossible because of
accidental fatal damage in the sample. However, we have managed to
obtain a rough estimate of $E_{out}$ from available PSTM images:
dashed rectangles in \fref{fig10}(b), (c) mark the areas over a
4-line average ($\approx 625 $nm) longitudinal cross-cut of each
image is obtained by means of WSxM software \cite{WSxMpaper}. The
resulting intensity profiles at \fref{fig10}(d), (e) show a clear
succession of steps, which we decide to characterize by the
arithmetic mean along the 5-micron central segment of each
plateau. Numerical estimates of $E_{out}$ are then calculated as
the ratio between $d_1,d_2$ and single slit values (see table at
the bottom left of \fref{fig10}). The coincidence of those
estimates with the calculated $E_{out}$ curve in \fref{fig10}(f)
is amazingly good, which encourages us to carry out conclusive
measurements in the near future. With respect to \fref{fig10}(f),
we finally have to remark that the radiative-to-SPP conversion
seems to be more efficient than its opposite, as far as $
E_{out}<2$ for any $d$.

\subsection{Telecom measurements}\label{tc}

\begin{figure}
  \flushright
  \includegraphics[width=0.84\textwidth]{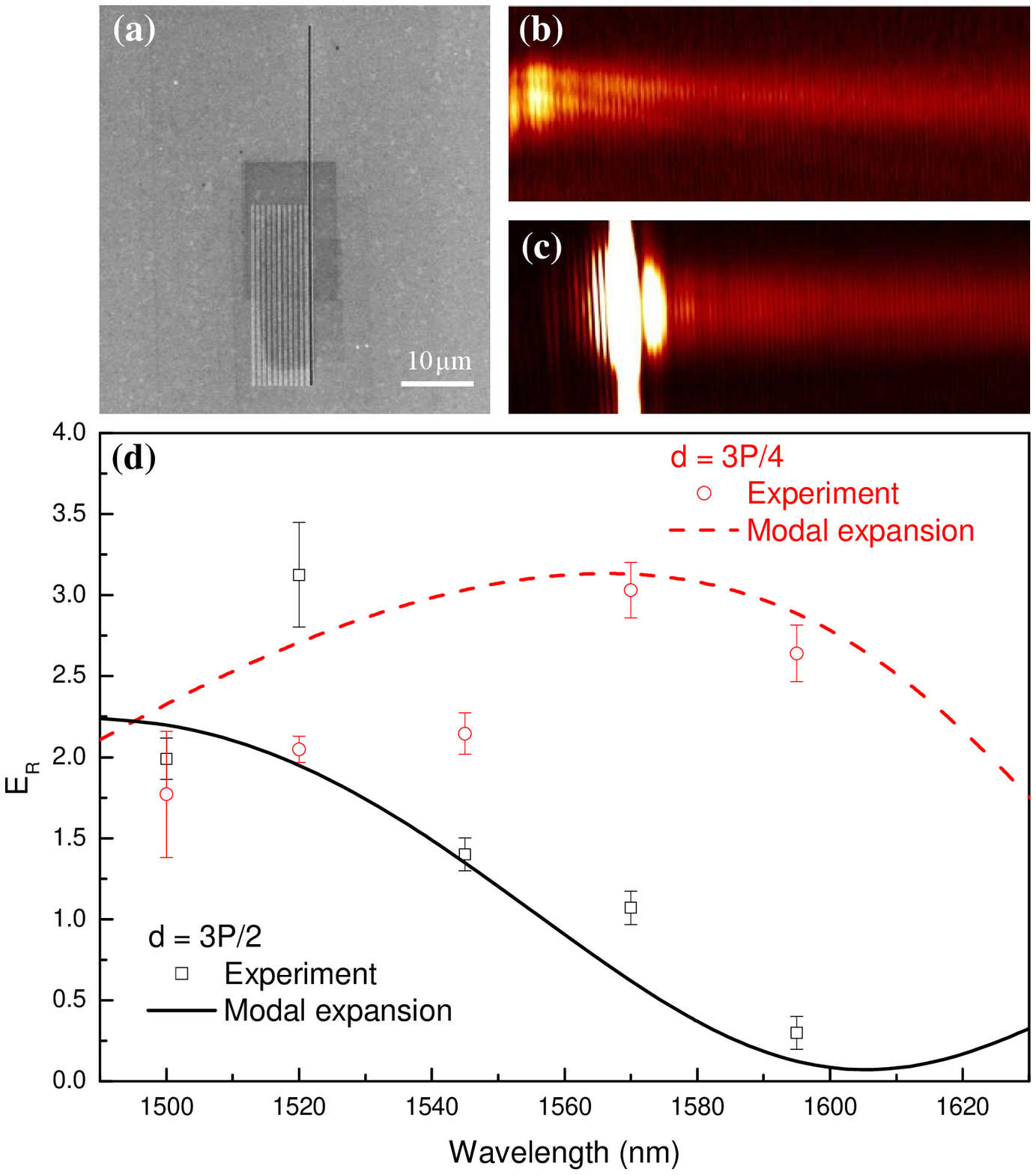}\\
  \caption{Spectral dependence of $E_R$ at the telecom range. \textbf{(a)}
  Scanning electron micrograph of the sample. The geometrical
  parameters are: slit length $L=50 \mu$m, slit width $a_0= 400$ nm,
  groove width $a=200$ nm, groove depth $w=100$ nm and array period $P=750$
  nm. \textbf{(b)} Near field image recorded with the laser beam focused at
  the ``isolated slit'' position of a sample with $d=3P/2=562$ nm. (Size = $70 \times 26 \mu m^2$, $\lambda=1520$ nm).
  \textbf{(c)} Same for slit+array focusing. \textbf{(d)}
  Spectral dependence of $E_R$ for slit-to-array distances of $d=3P/2=1125$ nm (experiment: squares;
  theory: solid line) and $d=3P/4=562$ nm (experiment: circles; theory : dashed line).}\label{fig11}
\end{figure}

Similar samples to those used in the NIR measurements were
designed to operate at the telecom range by up-scaling the period
of the array and its separation from the slit (see
\fref{fig11}(a)). However, in this wavelength regime, we found a
kind of instability in the illumination setup that resulted in a
noticeable variation of SPP intensity during the near-field scan
process, which takes about 45 minutes per image. As a consequence
of those intensity jumps, the technique used to evaluate the
``efficiency ratio'' in the NIR became unsuitable. Instead, we
found $E_R$ as the SPP signal ratio taken from each pair of
near-field images (with and without side grooves) at the same
distance from the slit, where its non-plasmonic field contribution
can be disregarded, whereas the SPP signal is still substantial
for the quantification ($\approx 50 \mu m$). To decrease the
uncertainty of thus obtained efficiency, a series of scans were
performed for every structure and wavelength measurements,
conducting independent adjustments, with the subsequent averaging
of the $E_R$ values obtained. Hence, the error of $E_R$ represents
a statistically estimated deviation.

A typical pair of near-field optical images is presented in
\fref{fig11}(b), (c). For telecom wavelengths, the SPP propagation
length is increased up to $\approx 200 \mu m$. Panel (c) features
a strong SPP beam propagating away from the slit in the direction
opposite to the array and thereby demonstrating unidirectional SPP
excitation. Averaged results and estimated errors for $E_R$
(previously reported in \cite{flt07np}) are rendered in
\fref{fig11}(d). Notice that the validity of our proposal is now
tested in a different way: for a given slit-to-array separation,
$E_R$ is measured within the wavelength range 1500-1620 nm, so
that the phase difference described by \eref{eqphi} is changed
with the increasing wavelength, providing the conditions for
constructive or destructive interference. Obviously, this spectral
dependence of the efficiency is different for different
slit-to-array separation, and we support that experimentally. For
the case of the sample with $d=P+P/2=1125$ nm, $E_R$ decreases as
the wavelength increases (with the only exception of a sharp peak
at 1520 nm), evolving from a favorable regime ($E_R \approx 2$) to
one in which coupling into SPPs is clearly diminished by the array
( $E_R <1$). Conversely, $E_R \approx 2$ for the sample with
$d=3P/4=562$ nm all over the range. As can be seen, the comparison
between experiments and  modal expansion calculation is rather
satisfactory.

\begin{figure}
  \flushright
  \includegraphics[width=0.84\textwidth]{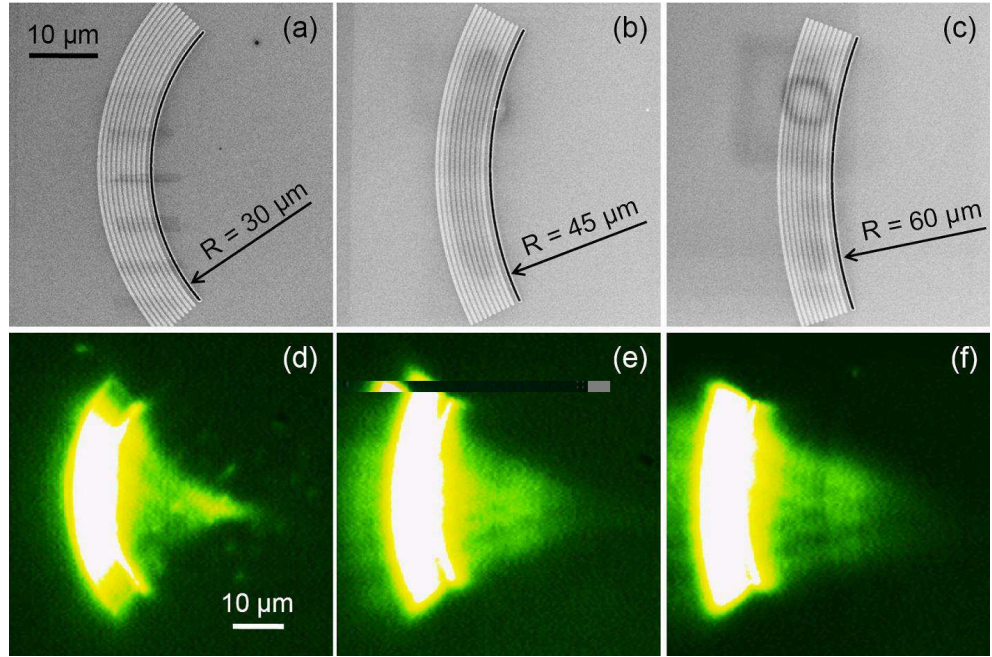}\\
  \caption{\textbf{(a)} Scanning electron micrograph of the curved
  structure, characterized by slit and groove widths of 400 and 200 nm,
  respectively, groove periodicity $P=750$ nm, groove depth $w=100$ nm and
  slit-groove distance $d=1125$ nm. Film thickness $h=280$ nm, curvature
  radius $R=30 \mu m$ and slit chord length $L= 40 \mu m$. \textbf{(b), (c)}
  Same for $R=45 \mu m$  and $R=60 \mu m$. \textbf{(d), (e), (f)} Far field images
  of SPPs excited on the structures(a), (b) and (c), respectively, recorded with a
  charge-coupled device camera.}\label{fig12}
\end{figure}

Finally, we have to mention that the proposed approach for the
excitation of localized unidirectional SPP beams can also be
combined with the appropriate design modifications to create
functional components for SPP focusing to a spot or tuning the SPP
beam divergence. If $E_R \ge 2$ is expected for a given slit+array
set, its circular bending may produce a converging gaussian beam
whose waist length and radius can be adjusted by means of the
curvature. Several curved SPP focusers has been previously
achieved
\cite{Nomura2005,Yin2005,Liu2005,Offerhous2005,Steele2006}, but we
find the mirror-blocked back-propagation to be a plus. Although
the rigorous modelling of SPP coupling at curved structures is
rather complicated and falls out of the scope of the present work,
we expect \eref{eqphi_R} to still provide a good estimation for
the proper design of the structure, at least as a starting point.
On that assumption, we have fabricated several samples consisting
of an arc-of-a-circle slit flanked by the corresponding array of
parallel bent grooves (see \fref{fig12} (a)-(c)). Geometrical
parameters ${a_0,a,w,P}$ are the same as in \fref{fig11}, whereas
slit-to-array distance is set to $d=3P/2=1125$ nm.

\begin{figure}
  \flushright
  \includegraphics[width=0.84\textwidth]{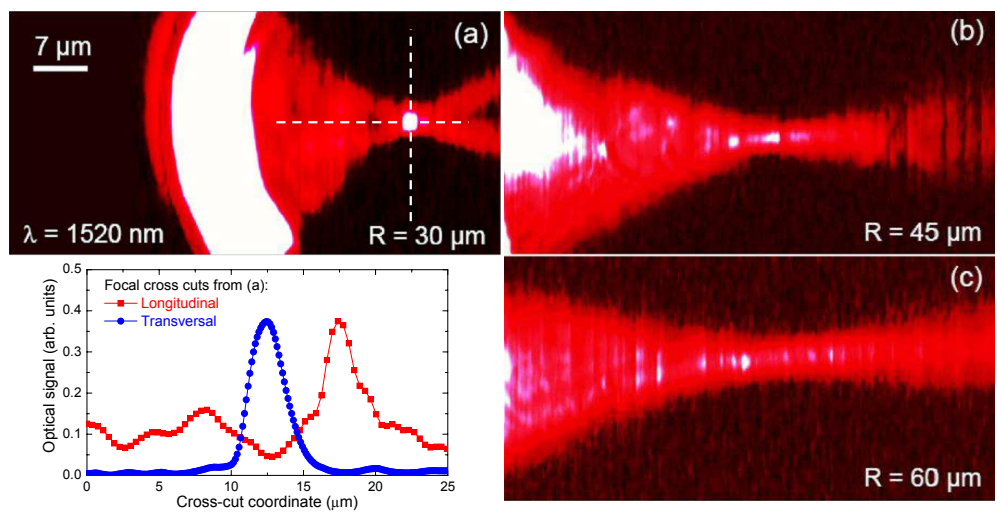}\\
  \caption{\textbf{(a), (b), (c)} Near-field images (size $64 \times 32 \mu m ^2$)
  of SPPs excited on the structures in \fref{fig12} at $\lambda=1520$ nm. Lower left panel depicts
  cross cuts obtained from (a) by dissecting the SPP focal spot along
  longitudinal and transversal directions.}\label{fig13}
\end{figure}

As shown in \fref{fig12}(d)-(f), the effect of SPP launching and
focusing can be appreciated already at the stage of far-field
adjustment due to weak out-of-plane SPP scattering by surface
roughness. Near-field images of SPP excitation on those structures
recorded at free-space wavelength of 1520 nm are presented in
\fref{fig13}. These images clearly demonstrate the property of a
curved slit to excite convergent SPP beam, with the effect being
sufficiently enhanced due to the side grooves (cf.
\cite{Yin2005,Liu2005}). With the smallest radius of curvature (30
$\mu m$), focusing to a confined spot having size $3\times 3$ $\mu
m^2$ is observed (see the cross cuts in the lower left panel of
\fref{fig13}). The SPP beams excited on the less curved structures
feature an extended waist (\fref{fig13}(b), (c)), which scales (at
least visually) according to expectations, providing a wider, and
hence less divergent, SPP beam. That might be useful for
particular applications, e.g. in sensing of elongated biological
samples or in coupling to low-numerical-aperture waveguides.

\section{Conclusions}
\label{conclu}

In conclusion, we have studied the SPP coupling-in at
sub-wavelength apertures with back-side illumination, presenting a
novel proposal for the modulation of such coupling-in by means of
a finite array of grooves. Our approach is based on a simple wave
interference model that, irrespective of the simplified
description of some of the physics involved, has been found in
good agreement with both sophisticated computer simulations and
experimental measurements at NIR and telecom ranges. We find this
to constitute a stimulating challenge for further developments on
a wide range of SPP devices.

\ack Financial support by the EU (Project FP6-2002-IST-1-507879)
and Spanish MEC (Project MAT2005-06608-C02-02) is gratefully
acknowledged. We thank J.-Y. Laluet for technical assistance.

\end{document}